\documentclass[11pt]{PoS}
\usepackage{graphicx}
\usepackage{amsmath}

\title{Twisted-mass reweighting for O(a) improved Wilson fermions}
\author{\speaker{Chuan Miao},$^{a,b}$
Harvey B.\ Meyer$^a$ and Hartmut Wittig$^{a,b}$ \\
\llap{$^a$}Institut f\"ur Kernphysik, Johannes Gutenberg Universit\"at Mainz, 55099 Mainz, Germany \\
\llap{$^b$}Helmholtz Institute Mainz, Johannes Gutenberg Universit\"at Mainz, 55099 Mainz, Germany\\
E-mail: \email{chuan@kph.uni-mainz.de}, \email{meyerh@kph.uni-mainz.de}, \email{wittig@kph.uni-mainz.de} }
\abstract{ 
We test the reweighting of the quark determinant of O(a) improved Wilson
fermions in the domain-decomposed hybrid Monte-Carlo algorithm.  Specifically,
we implement a reweighting in a twisted-mass parameter proposed by Palombi and
L\"uscher in $N_{\rm f}=2$ QCD. We find that at equal acceptance rate, the
algorithm is significantly more stable on a $32\times64^3$ lattice upon
switching on the reweighting parameter. At the same time, the reweighting
factor does not fluctuate strongly and hence is under control.  At equal
statistics, the uncertainty on the pion correlator is comparable to the case of
the standard, unreweighted algorithm.
}
\FullConference{XXIX International Symposium On Lattice Field Theory\\
	10-16 Jul 2011\\ Squaw Valley, Lake Tahoe, California}
\ShortTitle{Twisted-mass reweighting for O(a) improved Wilson fermions}

\begin{document}
\maketitle

\section{Introduction}

Simulating Wilson type fermions at small quark masses is challenging,
in particular because of the potential instabilities caused by the
fluctuations of the low modes of the Dirac operator. A direct study of
the low-lying spectrum of Wilson Dirac operators reveals that a
spectral gap forms for large volume
lattices~\cite{DelDebbio:2005qa}. However, occasional near-zero modes
may appear since chiral symmetry is explicitly broken by the
discretization. In simulations based on the hybrid Monte-Carlo
algorithm (HMC,~\cite{Duane:1987de}), this effect can lead to large
`spikes' in the history of the molecular dynamics Hamiltonian
violation. Some time ago, Palombi and L\"uscher 
proposed~\cite{Luscher:2008tw} to reweight the fermion determinant so
that the zero modes are suppressed by construction. The low-mode
contribution to any observable is faithfully restored by including the
reweighting factor in the ensemble average.  For a review of other
applications of reweighting with various fermion discretizations, see
the contribution of A.\ Hasenfratz at this conference.

In the simplest version of the idea, a twisted mass term is added to
the Dirac operator,
\begin{equation}
D(\mu) = D_W + i\mu\gamma_5 \,,
\label{eq:Diracmu}
\end{equation}
where $D_W$ is the $\mathcal{O}(a)$ improved Wilson Dirac operator
(see e.g.\ \cite{Luscher:1996sc}).  For $N_{\rm f}=2$, the weight
\begin{equation}
W = \det \left( \frac{D_W^\dagger D_W}{D_W^\dagger D_W + \mu^2} \right)
\end{equation}
should be included in the expectation value of the observable $\mathcal O$,
\begin{equation}\label{eq:reweight}
\left< \mathcal{O} \right> =\frac{ \left< \mathcal{O} W \right>_\mu}{\left<W\right>_\mu} \,,
\end{equation}
where $\left<\cdots\right>_\mu$ stands for the ensemble average 
for the modified Dirac operator $D(\mu)$. 

Evaluating the reweighting factor $W$ exactly is normally not
possible, nor is it in fact required; instead it can be calculated
stochastically.  One may add a set of $N$ pseudo-fermion fields
($\eta_k,\ k= 1,\dots,N$) to the theory with action
\begin{equation}
S_\eta = \sum_{k=1}^N (\eta_k,\eta_k) \,,
\end{equation}
and the reweighting factor is replaced by
\begin{equation}
W_N = \frac{1}{N}\sum_{k=1}^N\exp\left\{ \left(\eta_k, 
\left[1-\frac{D_W^\dagger D_W +\mu^2}{D_W^\dagger D_W}\right] 
\eta_k \right)\right\}.\label{eq:W}
\end{equation}
The simulation with respect to the modified Dirac operator proceeds as before
and the reweighting factor $W$ is estimated, for each gauge configuration,
according to Eq.\ (\ref{eq:W}) with $N$ randomly chosen pseudofermion fields.
Obviously, the larger the number of pseudofermion fields, the more accurate the
estimate.  Our tests show that 40 pseudofermion fields are sufficient to yield
satisfactory reweighted measurements of meson correlators on $32^3\times64$
lattices. As is visible from Eq.\ (\ref{eq:W}), fluctuations in the reweighting
factor $W$ that strongly suppress the contribution of certain configurations to
the statistical average can occur if the original Dirac operator $D_W$ admits
very small eigenvalues.

Convincing arguments were given by Palombi and L\"uscher that the
reweighting factor would not end up being exponentially small in the
volume, as one would at first expect. This follows from the fact that
the fluctuations of the lowest eigenvalues become smaller when the
volume is increased~\cite{DelDebbio:2005qa} (possibly with the
exception of a few).  Nevertheless an explicit test is needed to
ascertain that the twisted-mass reweighting idea works in practice,
and this is the subject of this work.

\section{Implementation}
We use the domain decomposition preconditioning \cite{Luscher:2005rx} of the
hybrid Monte-Carlo algorithm (DDHMC). Our simulation code is based on
L\"uscher\rq{s} open-source DDHMC code~\cite{CLScode}.  In this algorithm, the
whole lattice is divided into check-board coloured blocks. Using $\Omega$ or
$\Omega^*$ to denote the union of white or black blocks respectively, the Dirac
operator assumes the form
\begin{equation}
D_W = \left( \begin{array}{ll}
D_\Omega &  D_{\partial \Omega} \\
D_{\partial\Omega^*} &  D_{\Omega^*}\\
\end{array}
\right) \,,
\end{equation}
where $D_\Omega,\;D_{\Omega^*}$ denotes the Dirac operator respectively on block
$\Omega,\;\Omega^*$ with Dirichlet boundary condition, and
$D_{\partial \Omega}$ is the sum of all hopping terms from the
exterior boundary $\partial\Omega$ of $\Omega$ to the boundary
$\partial\Omega^*$ of $\Omega^*$. The fermion determinant is
factorized into local block parts and a global part. For $N_{\rm f}=2$ it reads
\begin{equation} \label{eq:fact1}
\det D_W^\dagger D_W = \det R^\dagger R\;\cdot\; \prod_\Lambda\; \det D^\dagger_\Lambda D_\Lambda\,,
\end{equation}
where $\Lambda$ runs through every block and 
\begin{equation}
R=1-P_{\partial\Omega^{*}}D_{\Omega}^{-1}D_{\partial\Omega}D_{\Omega^{*}}^{-1}D_{\partial\Omega^{*}}.
\end{equation}
With even-odd preconditioning on every block determinant, Eq.\ (\ref{eq:fact1}) can be further written as
\begin{equation} \label{eq:fact2}
\det R^{\dagger}R \cdot
\prod_{{\Lambda}}
\left[\det\left(Q_{ee}^{\dagger}Q_{ee}\right)\det\left(Q_{oo}^{\dagger}Q_{oo}\right)
\det\left(Q_{ee}^{-1}\widehat{Q}\right)^{\dagger}
\left(Q_{ee}^{-1}\widehat{Q}\right)\right] \,,
\end{equation}
where
  \begin{equation}
\widehat{Q}=Q_{ee}-Q_{eo}Q_{oo}^{-1}Q_{oe}
\end{equation}
and the site ordering has been chosen so that $D_\Lambda$ has the form
\begin{equation}
\gamma_5 D_\Lambda = \left(\begin{array}{cc}
Q_{ee} & Q_{eo} \\
Q_{oe} & Q_{oo}
\end{array}\right)\,.
\end{equation}
The forces of the MD evolution steps are
\begin{align}
(\omega,F_\Lambda)  & =
2\mbox{Re}\left(\widehat{Q}^{-1}Q_{ee}\phi_\Lambda,\delta_{\omega}
\left(\widehat{Q}^{-1}Q_{ee}\right)\phi_\Lambda\right)-2\mbox{Re}\mbox{Tr}
\left(\frac{\delta Q_{ee}}{Q_{ee}}+\frac{\delta Q_{oo}}{Q_{oo}}\right)\,, \\
(\omega,F_R)  & =
2\mbox{Re}\left(R^{-1}\phi_R,\delta R^{-1}\phi_R\right)\,,
\end{align}
where $\phi_\Lambda$ is the pseudofermion fields supported on the
even sites of block $\Lambda$ and $\phi_R$ is the pseudofermion field
residing on the block boundaries.

The modifications required to introduce the twisted mass term are relatively benign.
For the modified Dirac operator Eq.\ (\ref{eq:Diracmu}), the fermion determinant becomes
\begin{equation}
\det D(\mu)^\dagger D(\mu) = \det (D_W-i\mu\gamma_5)(D_W+i\mu\gamma_5)
\end{equation}
where the reweighting parameter $\mu$ appears with opposite sign for up and down quarks. 
The forces take the form
\begin{align}
\left(\omega,F_\Lambda \right) & = 
2\mbox{Re}\left(\widehat{Q}(\mu)^{-1}(Q_{ee}+i\mu)\phi_\Lambda,\delta_{\omega}
\left(\widehat{Q}(\mu)^{-1}(Q_{ee}+i\mu)\right)\phi_\Lambda\right)
\\ & \quad 
-2\mbox{Re}\mbox{Tr}\left(\frac{Q_{ee}\delta Q_{ee}}{Q_{ee}^{2}+\mu^{2}}
+\frac{Q_{oo}\delta Q_{oo}}{Q_{oo}^{2}+\mu^{2}}\right)\,,\nonumber \\
(\omega,F_R)  & =
2\mbox{Re}\left(R(\mu)^{-1}\phi_R,\delta_\omega R(\mu)^{-1}\phi_R\right)\,,
\end{align}
where
\begin{align}
\widehat{Q}(\mu) &=Q_{ee}+i\mu-Q_{eo}(Q_{oo}+i\mu)^{-1}Q_{oe} \,, \\
R^{-1}(\mu) &= 1-P_{\partial\Omega^{*}}D(\mu)^{-1}D_{\partial\Omega^{*}} \,.
\end{align}

Note that to calculate the global force $F_R$, one needs to solve the full
Dirac equation in every integration step. It is therefore important to solve
the equation efficiently.  The deflation
accelerated~\cite{Luscher:2007se,Luscher:2007es},
Schwartz-alternating-procedure (SAP,~\cite{Luscher:2003vf}) preconditioned GCR
algorithm is applied. Since the deflation subspace need not be exact, it is
constructed by a relaxation process, starting from a set of random quark fields
$\psi_l$, which are updated iteratively according to
\begin{equation}
\psi_l \rightarrow  `` D_W^{-1}{``} \;\psi_l \,,
\end{equation} until the condition
$
||D_W\psi_l|| \leq M\psi_l
$
is satisfied, where M is in the range of low eigenvalues of
$(D_W^\dagger D_W)^{1/2}$, and $`` D_W^{-1}{``}$ is an iterative procedure that 
approximates the inverse of $D_W$. In the current implementation~\cite{CLScode}, it 
is done with the SAP.  The operator $D_W$ is used in the relaxation 
procedure, the deflation efficiency for the operator $D(\mu)$ is equally good, because
the deflation subspace needs not to be exact and $\mu$ normally is small in practice.

The little Dirac operator, i.e. restriction of the Dirac operator $D(\mu)$
to the deflation subspace, is then specified as matrix
\begin{equation}
A_{kl}(\mu) = \left(\psi_k, D(\mu) \psi_l \right) .
\end{equation} 
The Dirac equation $D(\mu) \psi(x) =\eta(x) $ is separated into  two equations by acting with
projectors $P_L(\mu)$ and $1-P_L(\mu)$ from the left, where
\begin{equation}
P_L(\mu) \psi(x) = \psi(x) - \sum_{k,l=1}^N  D(\mu)\psi_k(x) A(\mu)^{-1}_{kl} (\psi_l, \psi).
\end{equation}

\section{Testing}

We have tested the reweighting method for two flavors of $\mathcal
O(a)$ improved Wilson fermions on large lattices ($24^3\times48$ and
$32^3\times64$) with fine lattice spacings (respectively 0.07 and 0.08
fm). The masses of the lightest pseudoscalar mesons in these ensembles
are approximately 360 and 300 MeV respectively.  Some details of the
simulation, for instance the molecular dynamics trajectory length
$\tau$, step size $\delta\tau$, the number of configurations
(ncfg) of the generated ensemble and acceptance rate of the HMC
simulations are listed in table \ref{tab:simu}.

\begin{figure}[t]
\begin{center}
\includegraphics[width=0.45\textwidth]{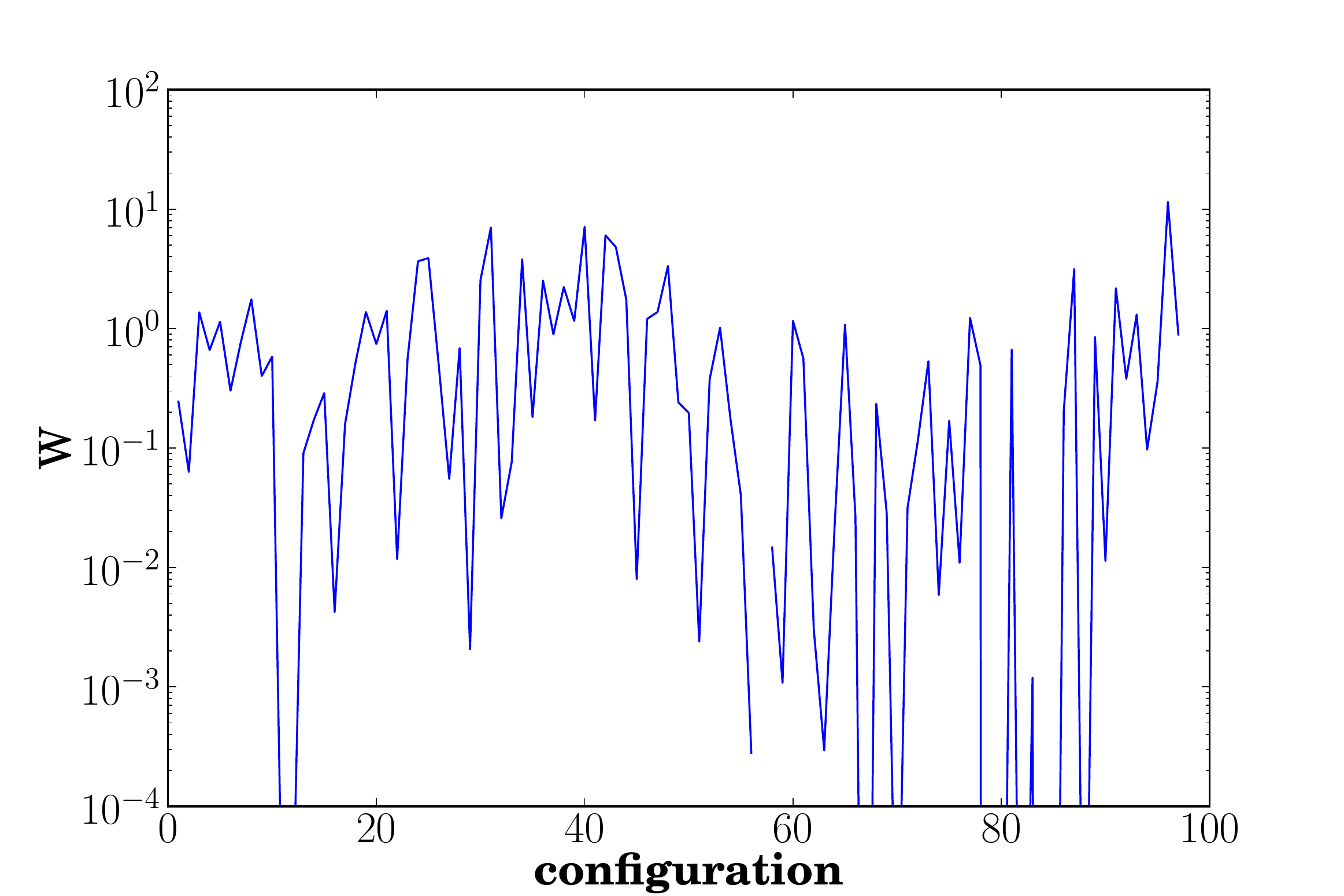}
\includegraphics[width=0.45\textwidth]{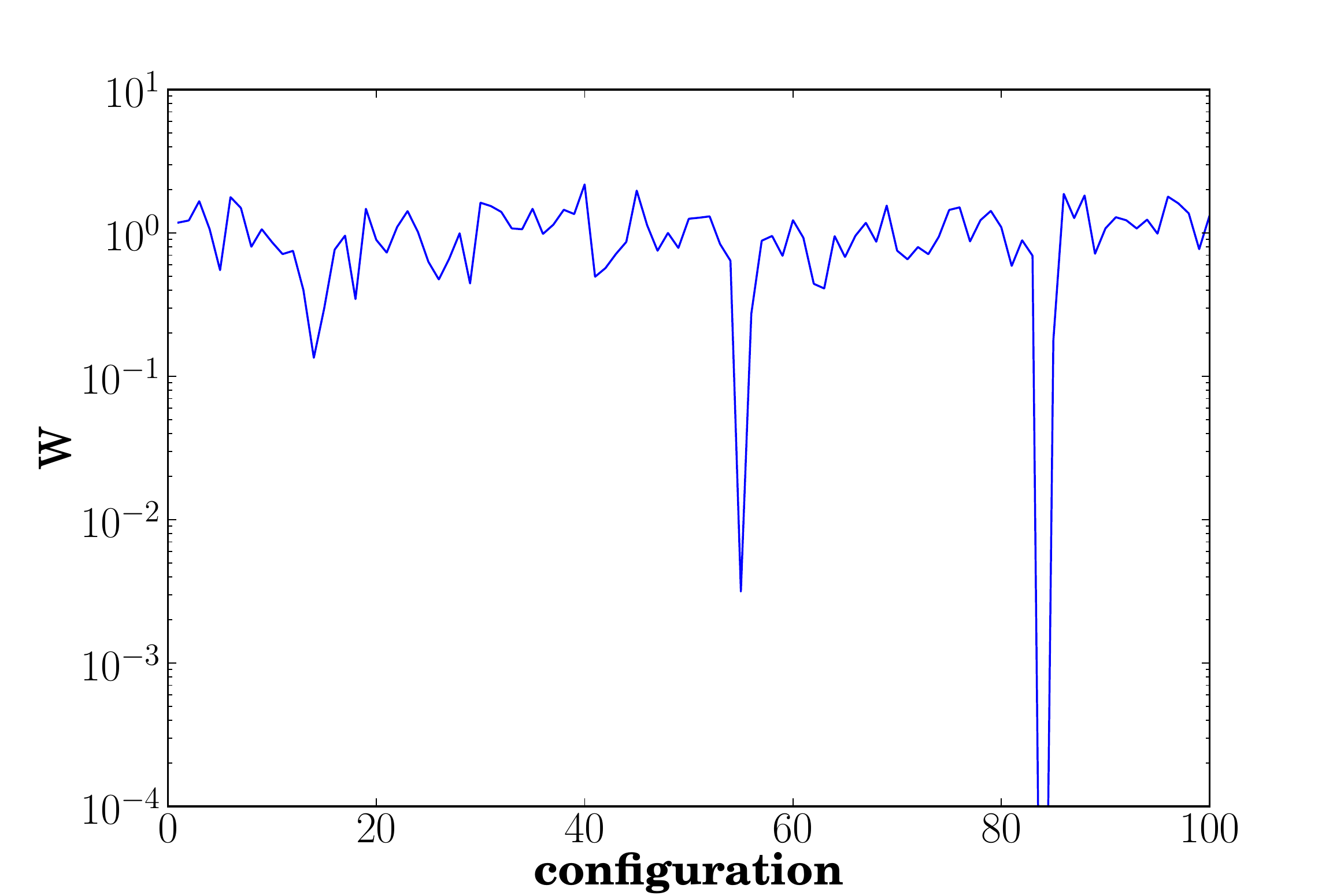}
\end{center}
\caption{Reweighting factor $W$ on $24^3\times48$ 
lattices for $a\mu=0.00569$ (left) and 0.003 (right).}
\label{fig:W}
\end{figure}

The reweighting parameter $\mu$ must be chosen with some care. Larger
$\mu$ accelerates the algorithm but leads to larger fluctuations in
the reweighting factor $W$. We have tested two values of $a\mu$,
0.00569 and 0.003, on $24^3\times48$ lattices.  The reweighting
factors $W$ are calculated according to Eq.\ (\ref{eq:W}).  Their
Monte-Carlo history is displayed in Fig.\ \ref{fig:W}, after being
renormalized so as to have an average value of one.

For $a\mu=0.00569$ (left plot),
the reweighting factor $W$ fluctuates strongly: about 20\%
configurations receive a weight that is smaller than 0.01. By contrast, for
$a\mu=0.003$ (right plot), the reweighting factor fluctuates moderately 
about the mean value.  We may use the kurtosis
\begin{equation}
K = \frac{\mu_4}{\sigma^4}-3 \,,
\end{equation}
where $\mu_4$ is the fourth central moment and $\sigma$ is the
standard deviation, to quantify the distribution of the reweighting
factor $W$.  For $a\mu=0.003$, the kurtosis is $K\simeq -0.11$,
showing that the fluctuations of $W$ is close to normal distribution;
while for $a\mu=0.00569$, kurtosis rises to $K \simeq 12.8$. Our 
choice of $\mu$ on the $32^3\times64$ lattice and the corresponding
kurtosis is also listed in table \ref{tab:simu}.

\begin{table}[b]
\begin{center}
\begin{tabular}{|c|c|c|c||c|c|c|c||c|c|}
\hline
Volume & $\beta$ & $a$/fm & $m_\pi$/MeV  & ncfg  & $\tau$ & $\delta \tau$ &  acc. rate & $a\mu$ & $K$ \\
\hline
$24^3\times48$ & 5.3 & 0.07 & 360 & 140 & 0.5 & 0.028 & 0.83 & 0.003  & -0.11 \\
\hline
$32^3\times64$ & 5.2 & 0.08 & 300  & 120 & 2 & 0.013 & 0.78 &0.001  & -0.37 \\
\hline
\end{tabular}
\end{center}
\caption{Parameters of the simulation, as explained in the text. The step size 
$\delta \tau$ is for updating the global force $F_R$.}
\label{tab:simu}
\end{table}

\begin{figure}[t]
\begin{center}
\includegraphics[width=0.45\textwidth]{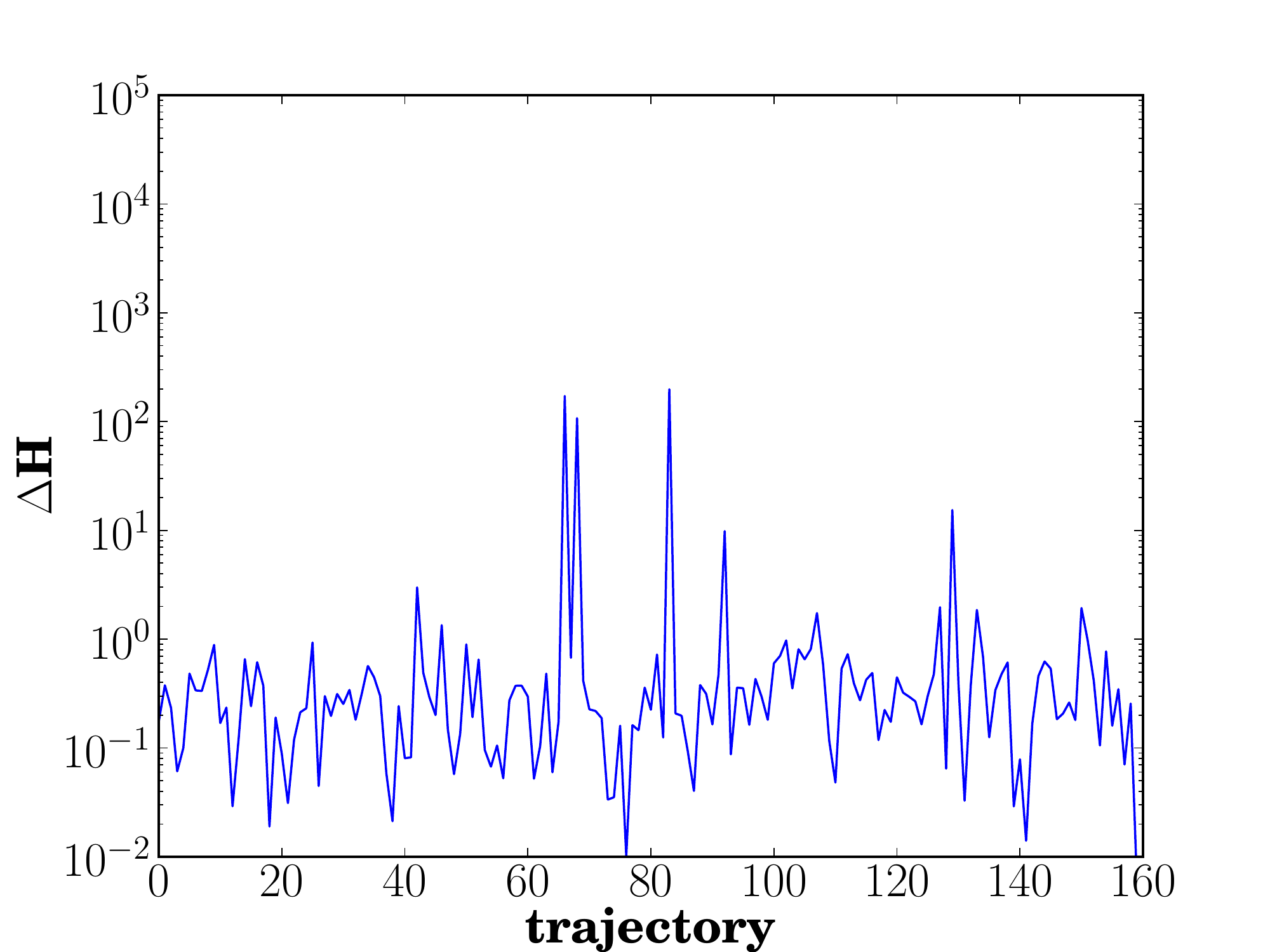}
\includegraphics[width=0.45\textwidth]{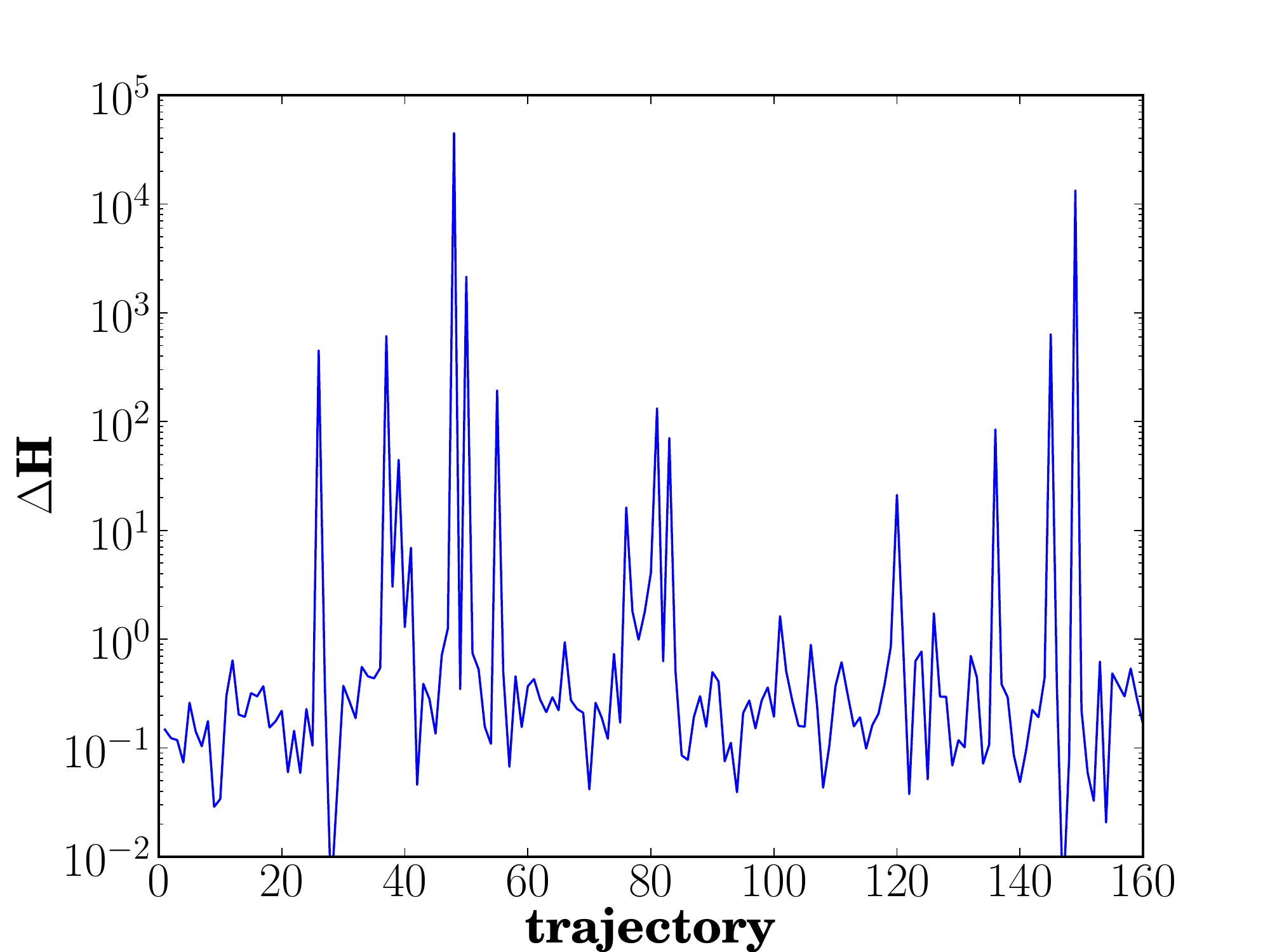}
\end{center}
\caption{History of Hamiltonian changes of molecular dynamics trajectories
 in HMC simulations on $32^3\times64$ lattices. The reweighting method
  (left) is more stable than the standard method (right). }
\label{fig:dH}
\end{figure}

In our simulations, we have tuned the integration step size of molecular
trajectories to achieve an acceptance rate of $\approx 80\%$.  For comparison,
we have simulated the same set of physical parameters with the standard method
($\mu=0$), using the same molecular trajectory length. We have tuned the
acceptance rate to be similar, $\sim$ 80\%. In order to do so, the integration
step size had to be roughly 10\% smaller. Although both simulations thus ran at
similar acceptance rates, we observe that the molecular dynamics trajectories
are more stable when reweighting is used compared with the standard method. We
plot the changes in Hamiltonian $\Delta H$ of molecular dynamics for each
trajectory in Fig.\ \ref{fig:dH} for simulations on the $32^3\times64$ lattice
(reweighting method on the left and standard method on the right).  For the
reweighting method, the history of $\Delta H$ is fairly stable, while for the
standard method spikes show up more frequently; the highest spikes are up to
three orders of magnitude higher compared to the reweighted algorithm.


\begin{figure}[t]
\begin{center}
\includegraphics[width=0.65\textwidth]{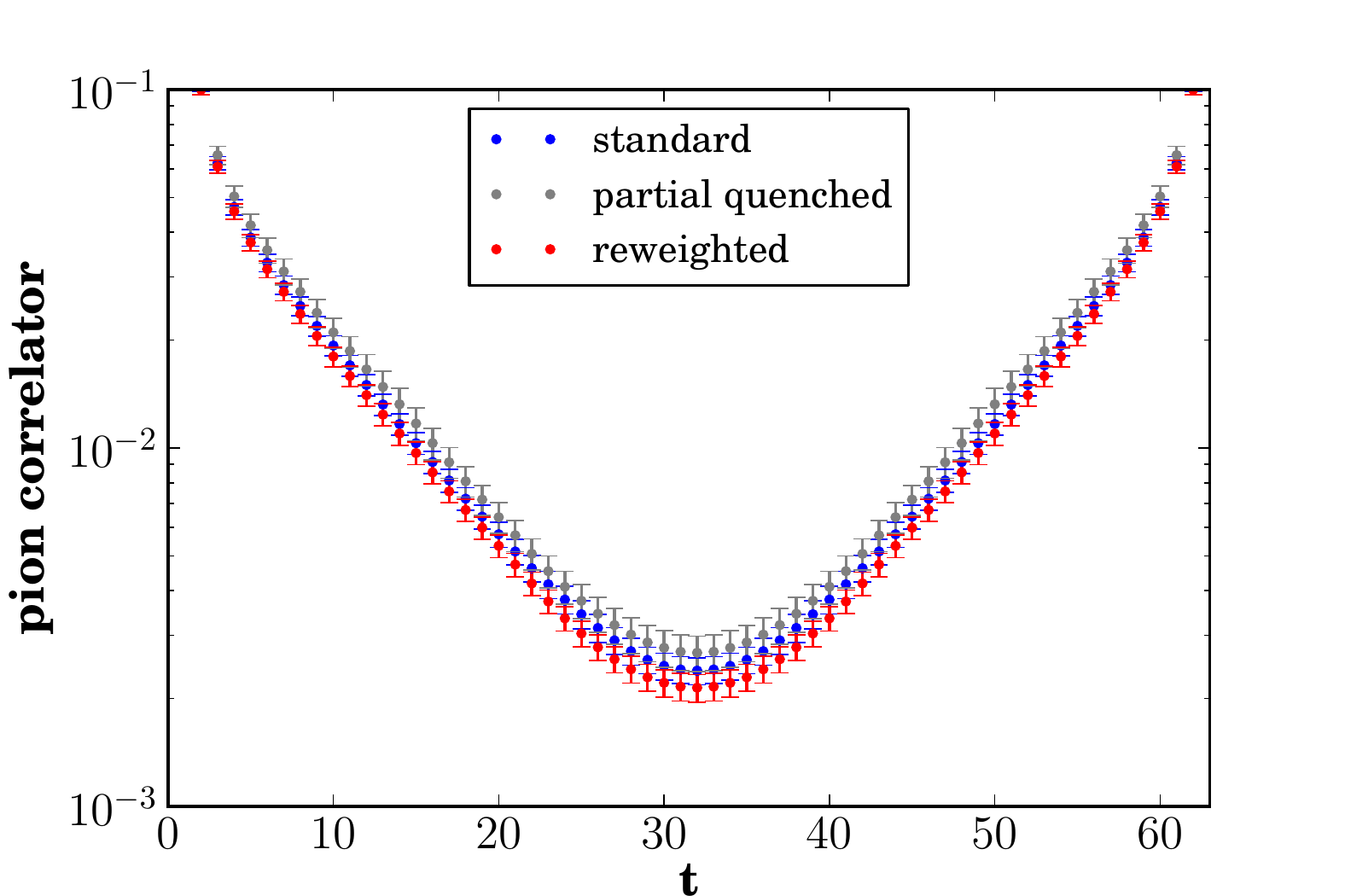}
\end{center}
\caption{Pion correlator on the  $32^3\times64$ lattice, with and without reweighting.
}
\label{fig:pioncorr}
\end{figure}

Using the reweighting method, we have calculated pion correlators and
compared them with the pion correlators measured on the ensembles
generated in the standard way. Practically one first measures the
correlator in the standard way on the ensemble generated with the
reweighting parameter $\mu$; we refer to this correlator as the
`partially quenched' one. Then one `corrects' it configuration by
configuration with the reweighting factor (\ref{eq:W}).  In
Fig. \ref{fig:pioncorr}, we plot the partially quenched pion correlator
(gray) and the reweighted one (red) and compare them with the
pion correlator obtained from the standard simulation (blue).
They are compatible within the uncertainties, and have comparable 
statistical errors. 

\section{Conclusion}

We have implemented a proposal by Palombi and L\"uscher to perform a
simulation with a quark determinant that is protected from the
fluctuations of the low-lying modes. We were able to find a
reweighting parameter for which the stability of the DDHMC algorithm
is significantly improved, and at the same time the reweighting factor
is under control.  As a test observable, we found that the correlator
of the pseudoscalar density is well behaved under the reweighting
(\ref{eq:reweight}). The behavior of other observables, such as the
nucleon correlator, remains to be explored.

In addition to studying the performance of the reweighted algorithm over more
than 1000 trajectories, it would be worth trying out other versions of
reweighting, where the net benefit in stability is potentially even larger.
Palombi and L\"uscher for instance made a proposal in this
direction~\cite{Luscher:2008tw} in which the UV modes are only affected at
order $\mu^4$.

\acknowledgments{We thank F.\ Palombi and M.\ L\"uscher for sharing part of
their code to compute the reweighting factor. We also thank our colleagues G.\
von Hippel, D.\ Djukanovic, B.\ Brandt and B.\ J\"ager in Mainz for helpful
discussions.  The work of HBM is supported by the \emph{Center for
Computational Sciences} in Mainz. The numerical simulations are performed on
the "Wilson" cluster at the Institute for Nuclear Physics of universit\"at
Mainz}

\bibliographystyle{JHEP}
\bibliography{/home/meyerh/CTPHOPPER/ctphopper-home/BIBLIO/viscobib}

\end{document}